# The role of open data in the transformation to Society 5.0: a resource or a tool for SDG-compliant Smart Living?

Anastasija Nikiforova[1,2], Miguel Angel Alor Flores[3] and Miltiadis D. Lytras [4]

**Abstract**–Open data are characterized by a number of economic, technological, innovative and social benefits. They are seen as a significant contributor to the city's transformation into Smart City. This is all the more so when the society is on the border of Society 5.0, i.e., shift from the information society to a super smart society or society of imagination takes place. However, the question constantly asked by open data experts is, what are the key factors to be met and satisfied in order to achieve promised benefits? The current trend of openness suggests that the principle of openness should be followed not only by data but also research, education, software, standard, hardware etc., it should become a philosophy to be followed at different levels, in different domains. This should ensure greater transparency, eliminating inequalities, promoting, and achieving sustainable development goals. Therefore, many agendas now have openness as a prerequisite. This chapter deals with concepts of open (government) data and Society 5.0 pointing to their common objectives, providing some success stories of open data use in smart cities or transformation of cities towards smart cities, mapping them to the features of the Society 5.0. We believe that this trend develops a new form of society, which we refer to as "open data-driven society". It forms a bridge from Society 4.0 to Society 5.0. This Chapter attempts to identify the role of openness in promoting human-centric Smart Society, Smart city, and Smart Living.

## 1 INTRODUCTION

Today, in the digital transformation- and smart city- driven world, citizens have become a fundamental part of the design, implementation, and governance of cities and, in particular, their infrastructure (Hernàndez, 2021). With the digitization and rapid development of emerging smart / intelligent technologies, more and more attention is paid to human- and citizen-cantered development.

    According to (Verhulst et al., 2021), emerging uses of technology can generate four forms of intelligence (4Is) - data intelligence, artificial intelligence, embodied intelligence, and collaborative intelligence. These forms of intelligence are able to improve the decision-making capacity of development practitioners by

enabling them to better understand or communicate relevant insights, as well as to accelerate progress towards the Sustainable Development Goals (SDGs). While some of them, such as Artificial Intelligence (AI), have already become buzzwords and are not anymore linked to one specific area only since becoming daily phenomena, there are some, which, although commonly used, are covered relatively rarely. This refers to the intelligence that Verhulst et al. (2021) call "collective intelligence" also known as "wisdom-of-crowd" (Suran et al., 2021). It is based on group interaction and improvements in communications, involving the use of technologies such as co-creation and crowdsourcing systems, smarter crowdsourcing tools, digital citizen assemblies, and open innovation platforms. The other 4Is components are embodied- and data- intelligence. Latest, however, is also closely linked with the subject of this Chapter since it relates to the use-cases reliant on datafication and analytics, involving the use of technologies such as open data and data collaboratives, in addition to others, such as the Internet of Things (IoT).

The openness of data is considered one of the crucial drivers for the sustainable economy and the knowledge-based economy in particular. It might have an impact on information and communication technology (ICT) innovation and creativity bridge in developing a new ecosystem in Industry 4.0 and Society 5.0 (Sołtysik-Piorunkiewicz & Zdonek, 2021; Nikiforova, 2021). This Society 5.0 is also known as the super-smart society and sometimes referred to as the "society of imagination" (Fukuyama et al., 2018), (Sołtysik-Piorunkiewicz et al., 2020) is expanding transparency and active participation in social issues by providing equal opportunities for all people and by integrating innovative technologies and society.

According to recent studies (Yoshida et al., 2021) in the recent movement towards smart societies, smart governments supported by GovTech and smart cities developed through Civic Tech have become known as dominant structures enabled by ICT. Both GovTech and Civic Tech share the common goal of giving citizens better and safer lives through their engagement with government and technology. Therefore, the development of online public services (e-services) is characterized using collaborative production methods involving various stakeholders and players. Open data plays a key role in it. Governments are the primary beneficiaries of so-called GovTech, where policies regarding OGD, and government transparency are developed to enable citizens' access to information and participation in government. This results in efficiency gains and saved costs. Civic Tech, however, encompasses a range of projects that use open government data to act in the public good (Yoshida et al., 2021).

What is more, the openness in both, data, science, technology (software or hardware) is considered as one of the keys for meeting Sustainable Development Goals (SDG), while supporting some of them "by default" simultaneously (general principles of open data covered by Open Data Charter applicable to all data to perceive and treat it as open data) and domain, which the open data represents. This was also emphasized at the 76th session of the United Nations General Assembly, highlighting that the openness contributes to the attainment of the United Nations Sustainable Development Goals (SDGs). This is also in line with Meschede & Siebenlist (2021), according to which open (urban) data promote smart and sustainable development in cities. They can also supply viable data for measuring progress toward the SDG.

Another point to be mentioned is that some challenges, especially those of global scale, such as climate change, pandemics and poverty are complex and there is little time to rectify them (Verhulst et al., 2021). This is also the case for crisis-management of different kinds, such as pandemics as COVID-19 we experience now, or other types of crises, including but not limited to economic, natural crises of the country or region. The level of complexity and especially short time slots within which solutions should or even must be developed makes open data an unprecedented artifact that allows individuals or groups of individuals to have an idea on how to combat the challenge to access the data without the need for their collection and preparation. This makes the possibility of the solution to be found to either a very tiny issue or even a very complex problem / crisis / disaster more likely compared to the closed data ecosystems, by which our past could be characterized. This is even more the case if the well-established environment for the above-mentioned co-creation is in place.

Of course, we do not pose the open (government) data as a "silver bullet" because, first, it should be available and accessible, but what is more important and crucial, a long list of prerequisites should be fulfilled. This list is related to both data usability, where the usability is a very multidimensional concept, including the trust in data, their quality, accuracy, timeliness, up-to-dateness, accessibility via API, their value that refers to the identification and opening of so-called "high-value datasets" etc. What is more, relevant ICT infrastructure from both technological and managerial perspectives with appropriate level of transparency is crucial, as well as the level of skills, knowledge and education forming both internet- and digital literacy should be met by individuals involved in both data supply and consumption. For the later, considering the specificities of open (government) data, their provision, use and reuse, specific term of "open data literacy" has been introduced (Weber et al., 2018; Loría-Solano & Raffaghelli, 2022).

What is more, open innovation typically considered as one of major benefits that open (government) data can bring, suppose additional requirements referring to the more complex concept of sustainable open data ecosystem (Linåker & Runeson, 2021; Lakomaa and Kallberg 2013; Janssen, Charalabidis, and Zuiderwijk 2012). This is also the case for smart open (government) data portals. It, in addition to above mentioned points, and considering actors dealing with the open (government) data at different levels, requires mechanisms for enabling cooperation and collaboration of different parties, thereby leading, and facilitating co-creation. The latter, i.e., enabling collaboration in a sustainable, collaboration- and cooperation- oriented open data ecosystem, is another very complex concept, which is not much studied, yet.

The objectives of this Chapter are two-fold: to define the Society 5.0 and OGD concepts and emphasize their interconnection, as well as to provide real-world examples proving these concepts are interconnected. The latter is expected to be achieved by providing a mapping of Society 5.0 determinants on the outputs of the OGD-driven Smart City examples. This also includes identification and elaboration on both determinants or prerequisites capable of promoting the development of the Society 5.0 by means of open data and barriers, which stakeholders of different types may face on the way towards sustainable smart city and super smart society.

The Chapter is therefore structured as follows: Section 2 defines the concepts of open data, open innovation, Society 5.0, Section 3 presents real-world use-cases and maps them on the features of Society 5.0, Section 4 establishes a discussion and concludes the Chapter.

## 2. MAIN CONCEPTS: OPEN (GOVERNMENT) DATA, OPEN INNOVATION, SOCIETY 5.0

### 2.1. Open data and open innovation

Open data, and open government data in particular, are considered a major enabler for transparency and trust, especially in the case of open government data and citizens' trust to the government, which is a pre-condition for the development of participative and smart community or city, and development at both social, economic and environmental levels. There is empirical evidence of the significant and positive direct relationship between open government data and institutional trust (Gonzálvez-Gallego et al., 2020), although there still exist areas for development. Open data are widely used for innovation, economic growth, competitiveness, job creation and societal progress (Francey and Mettler, 2021), as well as raising awareness about the events, including but not limited to COVID-19, spread of the disease, vaccination etc., forecasting, predicting, monitoring, tracking, contributing to data-driven decision-making and crisis management as well as the whole contemporary society.

Open data are more and more frequently used and transformed into value-added services and are considered driver for open innovation that is beneficial to society (Linåker and Runeson, 2021), (Smith and Sandberg, 2018), (Leviäkangas and Molarius, 2020). However, there is a lack of understanding and knowledge whether there exist real world examples, which prove open data-based services and solutions actually change / improve our lives and make our cities "smart"? This question was raised by Mainka et al. (2015) and is still valid. Unfortunately, there is a limited number of quantitative indicators of benefits of turning open data into (economic) value. However, there are a few interesting pieces of evidence found in the literature related to transportation, where probably two most expressive are (Leviäkangas and Molarius, 2020) and (Stone and Aravopoulou, 2018). They show that they can trigger the development of other derived services or those using the data provided, as well as provide direct benefits for both users and public authorities. We will elaborate on these examples and respective evidence in Section 3.

However, although the potential of open data is estimated to be very high and we will see it in the respective Section, there is a list of prerequisites / preconditions to be fulfilled in order to be able to gain those promised and expected benefits. Let us elaborate on the most widely discussed in more detail. Recent studies demonstrate that there are barriers / issues, which are emphasized in almost every study exploring this topic, which can potentially disrupt open data innovation development and open data initiative as a whole. Referring to the open innovation, Smith and Sandberg (2018), who aimed to explore how innovation barriers affect the use of OGD in different phases of the service lifecycle and how the perceptions of the barriers vary across different types of OGD users, have identified 38 barriers. Main barriers were further classified into 5 groups depending on the phase of service lifecycle: (1) strategy phase, (2) design phase, (3) transition phase, (4) operation phase, (5) continual improvement phase, where the overlapping of these phases is obviously permitted. More precisely:

- strategy phase associated with such barriers as lack of cooperation, understanding, domain knowledge, influencing and getting help from OGD providers and potential business partner;
- design phase is linked with lack of data, impeding data formats, poor documentation, poor data quality, and high task complexity, including lack of coding skills and knowledge of API use practices;
- transition phase described by lack of communication and support, slow data provision and issues with data maintenance, i.e., lack of up-to-date data, lack of time to develop services, and difficulties in developing services with unique value propositions and in gaining end-users, which is probably the phase which barriers are the least open data-related;
- operation phase characterized by a lack of communication from OGD providers, poor support, slow data provision and slow back-end reliability;
- (continual) improvement phase linked with the lack of understanding of how the services could be improved and difficulties in motivating efforts to do so.

Although it can be seen that many barriers relate to the overall digital literacy and lack of knowledge and or skills of developing innovative services, there are many barriers, which are developer-agnostic. In other words, while some barriers can be overcome either by seeking for appropriate team members or by acquiring required knowledge and skills, those open data-related are of more serious nature since can be understood as external and extrapolated on all potential stakeholders and innovative service makers.

This is something similar with what the most recent study (Linåker and Runeson, 2021) found, i.e., there is an expressed need for improved and more efficient feedback loops, collaboration, and a more demand-driven publication of OGD, i.e., rich data of high-quality, which refer us again to the concept of "high-value data" in addition to the need of establishing data quality as a default and mandatory pre-condition.

In addition to these activities, Mainka et al. (2015) point out that open data holders should also think about how to animate citizens, start-ups, and other stakeholders to reuse the data. They emphasize the role of workshops and hackathons organized with the goal to reuse open government data collectively to be then capable of doing so. This points out that it is not sufficient to provide data compliant with classical open data principles on the entry point such as national open data portal, and there is a highly expressed need for additional requirements to be fulfilled and continuous promotion activities in order to make open data initiatives successful, effective, efficient and sustainable. Furthermore, these prerequisites should also be managed and complemented if needed to be compliant with users' needs and current socio-economic and technological trends.

While the benefits of open data for data consumers are numerous, this is also the case for data producers and holders, who in many cases appreciate the benefits of the openness concept, especially "once-only principles". This was even more expressed in terms of COVID-19 pandemics. More precisely, the opening and publishing of data on the open data portal reduces the number of individual requests for information, thereby allowing resources to be rearranged to other tasks and to combat COVID-19 infection. This, indisputably, can be generalized and extrapolated on the open (government) data as a whole. However, this requires the data to be not only available and accessible, as well as of high quality, but also well-maintained. This requires additional resources to be spent since the data and all weaknesses associated with them become publicly available for the whole society. Therefore, in many cases there can be a resistance to open the data. Therefore, there are studies exploring challenges associated with the opening of these data, i.e., supply-side resistance to the OGD. As an example, Mainka et al. (2015) have presented challenges, by which can be characterized one of the most promising subsets of open data - open urban government data. They have classified them into: (1) political challenges mainly related to the fear of losing monopoly in public affairs, (2) legal challenges, where security, privacy, and copyright reasons are usually used to speculate, (3) governance challenges, (4) human resource challenges related to the above mentioned digital literacy of both data supplier and data consumer, (5) IT infrastructure challenges, (6) IT budget challenges related to the associated potential financial risk and different models of funding the open data initiative depending on the available budget.

Thus, although open (government) data can be characterized by a long list of potential benefits, unfortunately, its full potential has not been opened, yet due to some obstacles, which in many cases leads to the resistance from both parties, i.e., data supplier and data consumer, thereby delaying the development of Smart city and Society 5.0. As regards the latter, it is time to refer to the discussion on what this term stands for and how it relates to the concept of open data and how it affects Smart Cities and Smart Living.

## 2.2   Society 5.0: what was before?

Now, let us briefly define *Society 5.0*. First, let us briefly cover earlier "versions" of the society and determine what the current form is.

The first version of society, as we know it, is known as *Society 1.0*, or the "*hunting society*", which can be described as a society, where people coexisted with nature. This was followed by the *Society 2.0* or the "*agrarian society*", whose main activities were related to the development of irrigation techniques and the strict establishment of a settlement. Then there was perhaps a more noticeable paradigm shift with the invention of the industry, when an earlier version of the society evolved into so-called *Society 3.0* also known as the "*industrial society*", which is characterized and known for inventing a steam locomotive and launching mass production. Similarity and association with "*Industry x*" is not an accident, since the Society 3.0 has launched an *Industry* concept and movement that has evolved rapidly from the *Industry 1.0*, described mainly by mechanization and water and steam energy, to *Industry 2.0* associated with the beginning of mass production, assembly line and electricity. And then, with the invention of the computer, the subsequent distribution of data and information, the era of *Society 4.0*, also known as "*information society*", and *Industry 3.0* - computer and automation, have begun.

And now, we are talking about the next and most advanced version of society - *Society 5.0*, also called the *super smart society*. It is a new societal phenomenon, characterized as a highly human-centered society and even a society of the imagination. This form of society is characterized by the transformation of data and information into value by means of the imagination and creativity of different people or group of individuals, leading to human-centred data-driven solutions and value creation, usually by means of technological transformation and technological advances stemming from Industry 4.0 (Sołtysik-Piorunkiewicz, 2021; Federation, 2016). In other words, we use one of the most valuable and influential forms of a non-natural resource brought to us by *Society 4.0* - the data and information. This is expected to remain valid in the context of the new currently open and evolving paradigm of the Industry 5.0 posed by European Commission (2021) as the industry, which should facilitate the further development of the collaborative and co-creative vision, thereby complementing "*the existing Industry 4.0 paradigm by highlighting research and innovation as drivers for a transition to a sustainable, human-centric and resilient .... It moves focus from shareholder to*

*stakeholder value, with benefits for all concerned. Industry 5.0 attempts to capture the value of new technologies, providing prosperity beyond jobs and growth, while respecting planetary boundaries, and placing the wellbeing of the industry worker at the centre of the production process*". The question naturally stems from above - *whether we are already part of this society or rather are on our way to Society 5.0?* And in the latter case, *how it differs from country to country* and *what are prerequisites for determining the level of compliance with Society 5.0*. First, let us refer to the origin of the concept.

### *2.2.1 Society 5.0 origins*

The term Society 5.0 was approved by the Japanese Cabinet as a result of the 5th Science and Technology Basic Plan by the Council for Science, Technology and Innovation in early 2016 (Sphinx IT, 2019). The reason Japan was a country that has introduced the concept of the Society 5.0, although it is still unknown to many countries, relies on its highly developed economy, facing challenges including those, which are not typical for most countries, including natural disasters and the increasingly ageing population. It made a call for Japan to take action, and it came up with an ambitious societal – digital transformation plan.

They emphasize that the industry in its current stage does not organically transform societies, therefore there is a need for changes to be implemented in order to align the human with the industrial environment in an as natural way as possible.

The movement inspired other countries, their societies and economies and was the subject of discussion during the 2019 World Economic Forum. Yet despite 5 years already passed from the invention of this concept, most countries still are not aware of it. Most of them, though not familiar with the concept and its specifics, make at least some Society 5.0 compliant actions and policies. In order to better understand this, i.e., what these actions are, behaviour, policies, let us refer to some basics underpinning it. As the Information Society is a concept to which we are very familiar with, we will provide a comparison of these basics and emphasize the major changes. This should allow you to ask yourself *if you live in Society 4.0 or rather Society 5.0?*

### *2.2.2 Society 5.0: main features*

In order to better understand the concept of Society 5.0 and what it refers to, let us turn to the above-mentioned World Economic Forum 2019 and Keidanren (Japan Business Federation) first invented this concept (Nakanishi, 2019), (Keidanren, 2018)), which has introduced this concept and has provided us with the first set of features of Society 5.0, emphasizing the main differences with the previous form, i.e. Society 4.0.

Society 5.0 was presented as the "*Imagination Society, where digital transformation combines with the creativity of diverse people to bring about "problem solving" and "value creation" that lead us to sustainable development*". This concept should contribute to the achievement of the Sustainable Development Goals (SDGs) adopted by the United Nations, i.e., both sharing the same objectives. More precisely, it is characterized by 5 key areas: (1) *problem solving and value creation*, (2) *diversity*, (3) *decentralization*, (4) *resilience*, (5) *sustainability and environmental harmony*. All of them, being the core of Society 5.0, differ significantly from Society 4.0.

Firstly, when it comes to *problem solving and value creation*, according to which the course to "*a society, where value is created*" should be taken, there is a *liberation of the focus on efficiency* (economies of scale) established under the paradigm of Society 4.0 often referred to as plan-do-check-act (PDCA or Deming cycle). Instead, an emphasis is expected to be put on satisfying individual needs, solving problems, and creating value, which does not require the mandatory following of a uniform set of guidelines or standardized process.

*Diversity* refers to "*a society, where anyone can exercise diverse abilities*", which is not a "*uniformity*" anymore, but a liberation from suppression of individuality, including discrimination and alienation by ways of thinking and sense of values. This intends to make businesses more unique and diverse, i.e., it is expected that diverse individuals or groups of individuals will exercise diverse abilities to pursue diverse values in society. This should allow them to identify diverse tasks, challenges and needs in society and turn them into real business.

*Decentralization* by achieving "*a society, where anyone can get opportunities anytime, anywhere*" is an alternative to the previous concentration achieved by means of a liberation from disparity. This means that anyone will be able to get opportunities to participate at any time and any place. It must be pointed out here that, although a few years ago before pandemic, it would be difficult to imagine in on a global scale, where only a few players were ready for this, the pandemic and restriction that were assigned to us during the COVID-19 have shown that it can be done in at least a certain sense, i.e., by means of remote work. While many sectors have made clear that this model is appropriate and should be followed even after the pandemic, in some cases there is still strong resistance to this model. However, it is clear that the remote or a hybrid working mode serves only as one small example of this principle, as this principle applies to other forms such as participation in processes and data sharing with others, perhaps smaller companies, or educational and working opportunities provided to people born in poverty or remote areas.

*Resilience* (especially against terrorism and disasters in physical spaces and attacks in cyberspace) and sustainable development standing for "*a society, where people can live and pursue challenges in security*" aimed at ensuring liberation from anxiety and vulnerability. This is intended to be achieved by means of a new, diversified, and decentralized social infrastructure.

*Sustainability and environmental harmony* which refers to "*a society, where humankind lives in harmony with nature*", which is understood as a liberation of resources and environmental constraints that we have experienced in recent decades due to the mass consumption of resources, with a major impact on the environment.

Society 5.0 aims to address and resolve the critical social issues facing humankind today from poverty to clean energy that contributes significantly to the achievement of the SDGs. Thus, both the Society 5.0 and the United Nations SDGs are considered to be sharing common objectives and being interrelated (for a broader discussion on how the Society 5.0 expects to meet SDGs, see (Keidanren, 2018)).

## 2.3   Society 5.0: the role of open data

Now, let us emphasize the compliance of the open (government) data objectives with those forming the Society 5.0, thereby inviting to treat the open data as the source and even a tool for achieving and maintaining Society 5.0. Although this compliance and overlapping of main principles of the concepts mentioned above are somewhat obvious, current literature presents only several studies acknowledging this, more precisely (Sołtysik-Piorunkiewicz et al., 2020) and (Nikiforova, 2021).

Sołtysik-Piorunkiewicz et al. (2020) were the first, who found that the concept of "open data" is closely linked with the Society 5.0 term. Their findings were based on the results of the bigram-centred analysis, where common concepts and techniques (including machine learning (ML), artificial intelligence (AI), predictive models, visualizations, interactive maps etc.) have been identified and mapped one onto another, identifying them as artifacts shared by both concepts. As regards types of data used, the authors identified that geodata/ geospatial data are the most popular type of data, followed by transportation-related and health-related (open) data. This study, however, was inspired by Zuiderwijk et al. (2015), in which "open data performance expectancy" involved reflecting the benefits of using open data, which positively influences behavioural intention to adopt open data technologies. Sołtysik-Piorunkiewicz et al. (2020) found that the co-creation of a value for the sustainable ecosystem is in creating a bridge between Industry 4.0 and Society 5.0, where open government data play a role being a resource and a tool at the same time.

The authors found that the main course of presenting the benefits of open data services is to promote them as tools to provide real-time data and information on public issues in areas such as transport, health, economics and finance, education, culture, and sport. Based on their bigram's analysis, "real-time" was found to be the most popular in the description of open data-based services, followed by bigrams related to specific business sectors dominated by transport, location, traffic, and parking, which could be associated with Internet of Things (IoT) and more specifically Internet of Transportation and Internet of Vehicles, smart transportation. Surprisingly, geospatial data, while found popular, was significantly less popular, as might be expected, the service description cites 13 times less than "real time". However, analysis of the used data types revealed that geospatial data was one of the most popular types of data to be used in developing these services. Geodata/ geospatial data is thus the most popular type of data, followed by transport-related and health-related open data. As in (Nikiforova, 2021), we believe that today, in the light of COVID-19, health-related data and pandemic-related OGD and their further reuse would be one of the most popular data domains.

Taking a step back to the fact that only several studies have previously acknowledged the common objectives of both, this is something similar to what we have referred to before, i.e., although the concept of Society 5.0 may be unknown for many societies and governments, the principles by which Society 5.0 is characterized are sometimes the focus for their agendas or even already in place.

More precisely, they can be divided into those principles of Society 5.0, which can be mapped onto open data principles, and the aim / objective of the open (government) data. As regards the principles of open data, they are generally 8 (Sunlight Foundation, 2007), i.e., data may be considered as open data if they are: complete, primary, timely, accessible, machine-readable / machine-processable, non-discriminatory, non-proprietary, licence-free. This means that the data should be as complete as possible, available to everyone without any restriction on gender/sex, race, religion, citizenship etc. without registration requirements and copyright, made available as soon as possible - as they are gathered and collected - in order to maintain their value, reflecting all those recorded about a particular subject, and in a format on which no entity has exclusive control.

This means that open data by themselves, i.e., by definition, fulfil requirements *diversity* and *decentralization*. *Problem solving and value creation*, as well as *resilience, sustainability, and environmental harmony*, however, are objectives to be met by the reuse of open data - their transformation into the value facilitating innovation and promoting societal progress.

For this purpose, an extensive research is made over the decades to identify key factors affecting users' willingness to use them and providing them with the opportunity of doing so by both technological advances by which the open data portals get supplied and social activities such as webinars, hackathons etc. aiming at raising awareness of data, how they can be reused and transformed into the value and determining the areas, in which the solution are actually needed. Here, as an example, hackathons devoted to smart city or crisis management (not only pandemic but also natural disasters) topics are becoming increasingly popular. They, however, in some cases, are then turned into start-ups, which might be also in line with the above-mentioned diversity and unique approach to be followed depending on the task solved by the individual or a group of individuals.

In other words, open data and OGD in particular, share the same principles as Society 5.0 with some of them lying in the definition and prerequisites to be fulfilled by data to be considered as open data, while others are the objectives of OGD availability.

To prove the latter on more specific examples, we have to refer to real-life examples and examine whether and how they facilitate "sustainability and environmental harmony" and "resilience".

## 3. USE-CASES

In this section we refer to the real-world use-cases, which were selected based on several factors forming an inclusion criterion. First, they should mandatory utilize open (government) data as either the main source of the data or as a secondary or auxiliary source of data. In addition, they may produce open data. Second, in order to make an intersection and close relationships between both concepts of the open data and Society 5.0 more visible, we seek for the use-cases that represent Society 5.0 focus areas. They are cities and regions, energy, disaster prevention, healthcare, agriculture and food, logistics, manufacturing and services, finance, public services (Mavrodieva et al., 2020). Considering the definition of these areas and the nature of the concepts we refer to, we do not treat manufacturing and services as an independent category, being overlapping with other areas that will be demonstrated in the following subsections. Thirdly, which partly overlaps with the previous one, considering the fact that the open data are often discussed in the context of the SDG and their support, we were seeking for the examples/ showcases, which would cover different SDGs, thereby extending the scope of the areas in which these showcases appear.

First, we discuss the use-case selected. We cover use-cases, with which we are either familiar with, by being involved in their development and involvement or by being familiar with people who have been involved in it or have access to sufficient documentation of these showcases. In some cases, very prospective but less widely discussed use-cases have been selected to emphasize more promising examples or areas by moving towards Society 5.0 and Industry 5.0. Then, we map their characteristics on the main features of Society 5.0 we have discussed in the previous Chapter.

Table I provides a summary of these use-cases providing both the focus area, country in which it has been developed, organization and project within which it has been developed and a brief description of the case.

**Table 3.1**. Summary of use-cases by focus area

| Focus area | Policies (defined in (Mavrodieva et al., 2020)) | Country | Organization and Project | Case |
|---|---|---|---|---|
| Energy | "Development of affordable sustainable energy; development of micro-grid systems to respond to local conditions" | Portugal | Energias de Portugal (EDP) and Opendatasoft, EDP's Open Data Portal https://opendata.edp.com/pages/showcase | Energias de Portugal Open Data Portal provides resources about the energy sector in Portugal for spurring community innovation and confronting energy industry challenges |
| Disaster Prevention | "information sharing across organizations; utilization of digital technologies; continuation of medical services and aid in the event of disasters" | Indonesia | Yayasan Peta Bencana https://info.petabencana.id/ | Real-time disaster information sharing system of digital technologies; continuation of medical services and aid in the event of disasters |
| Healthcare | "Focus on prevention and individualized healthcare services;access to personalized life-stage data, utilizing AI-based medical services, such as telemedicine" | Uruguay | DATA ATuServicio http://atuservicio.uy/ | Access to personalized healthcare providers data utilizing an open-data-based platform for comparison and evidence-based decision-making |
| Cities and regions | "Improved data sharing on energy, transportation, water, waste, human traffic, etc.; community decentralization in suburban and rural areas; respect for diversity" | United Kingdom | Transport for London (TfL) https://tfl.gov.uk/ | All public TfL data (or 'open data') is freely released for their further re-use. Users and software developers in particular are encouraged to use these feeds to present customer travel information in innovative ways - providing they adhere to the transport data terms and conditions. |

| Agriculture and food | "Utilization of technology for crop growth and optimization of the food value chain; inclusion of various actors" | Canada | Global Open Data for Agriculture and Nutrition (GODAN) | By making open data on agriculture & nutrition available, accessible & usable for all, GODAN tackles extreme poverty, eradicates hunger, improves nutrition, and achieves food security. |
|---|---|---|---|---|
| Logistics | "Utilization of technology for automation of logistics; data sharing across the whole supply chain; personalized products responding to specific customer needs" | Germany and Japan | FIWARE Foundation Member TIS Inc. and the University of Aizu https://www.fiware.org | FIWARE-based automated logistics system using heterogeneous mobile robots enabling the management of real-time data from various robots, IoT sensors and open data to control and manage robots and their operations. |
| Public services | "Improved services by public administration based on digitization and improved data sharing; establishing safety nets in response to safety issues" | Buenos Aires<br><br>Spain | BA Obras, https://www.buenosaires.gob.ar/baobras<br><br>Papelea https://www.papelea.com/ | BA Obras, where the most up-to-date information on public works to be useful for citizens is provided.<br><br>Papelea provides the support for citizens of Spain and Mexico and solves their questions of both legal and administrative nature by collecting public information from various websites of these countries and providing it in a summarized and structured manner. |
| Finance | "Diversification of financial services with the help of digital technologies; better distribution of funds across society; improved access to financial services, based on utilization of crypto-currencies and token economies, such as blockchain" | Spain | BBVA API market in Spain https://www.bbvaapimarket.com/en/ | Informed consent-based open banking, enabling generation of new business models, creation of global and local alliances providing added value through APIs, offer improved service compliant with the users' needs |
| Manufacturing and services | "Focus on services, not hardware; customers will be able to order items specifically designed for their needs; support for small businesses to produce high-quality goods" | see other areas | see other areas | see other areas |

Let us provide a brief overview of these use-cases, by covering their main idea and objective, area, challenge they aimed to combat and what role the open data play in them.

### 3.1 Energy area

Renewable energy is one of the most ongoing topics today being of interest for both private and public authorities and the society as a whole, as well as representing the 7th Sustainable Development Goal of Affordable and Clean Energy. Therefore, the first use-case we refer to is the Energias de Portugal (EDP) - one of the main producers of wind energy in the world (European Commission, 2021), and the respective EDP

Open Data Portal, which was launched to provide resources related to the energy sector (in Portugal) for spurring community innovation and confronting energy industry challenges. According to (OpenDataSoft, 2018), in a number of cases, they were able to meet 100% of Portugal's electricity demand from renewable energy sources alone that is usually claimed to be unattainable, thereby providing a counterexample. As a result, being considered as a marvel in this area, they have made a commitment to energy innovation and sustainability through collaboration and information sharing by means of open data.

*Challenge and Value*

Energias de Portugal deals with the new challenge of the high penetration of renewable energy - intermittency. The production of these energy is based on renewable energy sources, which are not mobile, i.e., non-dispatchable, where energy / power managers are unable to control the wind or the sun as the source of these energy, making them less manageable compared to the conventional and polluting fuels. Today's electricity / power grid, as explained by Robert Fares (2015), was "conceived around the ability to control the generation of power to produce the right amount of electricity at the right time to meet demand, with limited storage capacity". As a result, energy system operators are looking for leading technologies to promote and even facilitate the global distribution of renewable energy and to improve the operation, management, and reliability of respective assets.

Open data are considered to be able to play an important and even decisive role here as is considered not only by us, but also by EDP. Therefore, EDP has transformed this task and even challenge into an opportunity to establish a community to create new services in a collaborative manner. Their open data initiative is aimed at providing resources for communities of academics, researchers, and other stakeholders, facilitating co-creation. In addition to the opportunity of contributing to community innovation in order to address energy related challenges, which EDP calls "challenges of tomorrow", they believe that sharing the data will benefit the society as a whole by informing them, i.e., creating the knowledge of the energy sector, current state, room for improvements and challenges to overcome.

To this end, EDP has turned to an open innovation approach, establishing closer relations with customers and partners launching a respective data portal, where the first input came from the Hack the Wind hackathon, seeking for new solutions to the operational challenges affecting wind turbines, held as part of the Wind Europe event - the wind energy sector's largest conference. EDP believes that by providing access to these data, users are given the opportunity to freely define their showcases / use-cases / re-uses and share these solutions to the key problems facing the energy sector (according to their view), thereby exploiting the collective intelligence of their communities. In this way EDP aimed to respond to the demand of data from the technology community, provide data for start-up projects, academic and scientific purposes (since open innovation approaches lead to new approaches to industrial challenges), thereby facilitating and encouraging the extraction of the value from these data and development of new services and be part of the ongoing transformation of the energy sector.

However, although the general idea seems to be very promising and in line with the current trends, it should be admitted that the respective portal has only 13 datasets, where the most recent datasets are modified in 2019, while the data that could be potentially used answering the current challenges are on 2017, i.e. Sunlab Faro weather station data and Production data and Temperature of the photovoltaic modules, which reduces the value of these data significantly, especially given that these data cover only one region - Faro. This is also something similar to reuses available, where the latest reuse is of 2018.

Although it might be stated that this data portal is not very active anymore and there is no evidence that the data published was actually reused by the communities and the value has been derived by transforming the data in solutions, there is a thread of informative materials, which provide a sufficient background on the current state of the art and challenges to be overcome in this sector. In other words, although the original aim of the portal since that has not been successful, i.e., users are not sharing their reuses, the data are not constantly opened and maintained, it partly meets their expectations and provides the visitors with the insights on the subject.

## 3.2 Disaster Prevention area

The topic of disaster prevention, where disasters can be both human-made and natural (droughts, floods, landslides, volcanic eruptions, and earthquakes) affecting people wellbeing and even lives as well as the economics, is another area where the value of open data is considered very high and have already proved themself to be as such. It is clear that the open data is not the only source that is important in this respect, where social media has previously been assessed as a valuable resource for extracting, collecting and tracking data on the ongoing emergency or the first signs of it. However, the value of open data is in the possibility of using it as a source for different types of models and forecasts and determining potential disasters before they occur and preparing or even preventing them.

This is all the more so for Japan, a country where the beginning of the open data movement can be found after one of such disasters, i.e., the East Japan Earthquake and Tsunami in 2011, after which data that may be potentially related to disaster prevention were recognised as the most valuable data for disclosure (Kanbara et al., 2022). Later in 2015, a Sendai Framework for Disaster Risk Management has been defined, according to which disaster risk management policies and practices should promote real-time access to reliable data, the use of space- and ground- based data, including geographic information systems (GIS), and the use of innovative technologies, including but not limited to AI and ML, improving not only data collection and supply, but also their processing, analysis and derivation of the value from them. A similar conclusion was made as a result of the event organized by the United Nations Economic Commission for Africa "The Use of ICT for Disaster Risk Management and Climate Change Mitigation" in 2013, pointing to the need for governments to raise awareness on disaster risk management and climate change mitigation and to encourage countries to include disaster risk reduction in national policies by introducing policies on open data on accessibility and usability of disaster information (UNECA, 2013).

The study conducted by Kanbara et al. (2022) have stressed that, even if financial resources are limited or not available at all, the publishing of data consistent with the open data principles on such daily aspects as traffic, public works and public health and geospatial data and information, in particular, can save lives in crisis situation and can be used as an effective tool of information communication and rapid response by means of a coordinated approach, i.e. "co-delivering" risk communication, coordination, strategy setting and decision-making before, during and after a disaster.

Another example is an Indonesian Petabencana.id CogniCity Open-Source Software-based platform for emergency response and disaster management in megacities in South and Southeast Asia (Widyanarko, 2018).

*Challenge and Value*

Petabencana.id is a digital information-sharing ecosystem or disaster management service, which produces real-time disaster maps by using the data gathered/ crowdsourced from social media and open data repositories (the developer refers to them as government agency validations) of Jakarta. The data used by it is the data, by which citizens describe the current situations in the social media, including selfies, and since (if) these posts and tweets in particular, are geo-referenced, open-source software CogniCity OSS used by this service map the tweets and share them through a publicly available map. These data are then sorted, analysed and displayed if the risk has been determined and confirmed in real-time, thereby providing citizens, communities, and government agencies with the critical information (Widyanarko, 2018). In this way the map operates as decision-making support for citizens to evacuate and navigate the city during floods and for government agencies to act in response to life-threatening situations.

The developers of this service describe the paradigm they are using as "people are the best sensors", by collecting the data directly from the users "at street level", which allows to remove expensive and time-consuming data collection and processing activities (although not all of them since these data also suppose processing and sorting, as well as extracting relevant data). Another prerequisite for the success is understanding and satisfaction of both prerequisites of transparency and collaboration, which are considered to be a key for success and accurate and timely determination and response to the risks. This showcase has also led the developers to the conclusion that the combination of open-source software and open data support collaborative efforts for adapting to climate change, thereby allowing stakeholders to review and inspect both the service and the current situation and then develop complementary tools that could further enhance resilience.

The method used, i.e., crowdsourcing, provides a competitive response to the limitation of data collection during a disaster situation, when conventional data collection and management of sensor-produced data limit the collection of real-time and street-level data, which is crucial for emergency response. Conventional sensors must be located in static places and require regular maintenance to enable real-time data to be provided, which should be also sufficiently accurate, while sensor data are sometimes described as noisy, and data collection is limited to the number of sensors available in the city, while stakeholders need to get an comprehensive overview of the emergency in the short time.

According to the PetaBencana.id, this platform has been used by millions of resident users since its announcement in 2013, being thereby able to take time-critical safety and navigation decisions in case of emergency disasters. What is more, it was adopted by the National Emergency Management Agency (BNPB) to monitor floods, improve response times, and exchange critical information on emergencies with citizens, allowing better information exchange and data coordination between citizens and government agencies, promoting fair and collaborative resilience to climate change. The developers believe that community-driven data collection, sharing and visualization reduces flood risk and helps in rescue efforts. As a result of this project, it was recommended as a model for community involvement in disaster response and best practice for disaster information collective resources by the International Federation of the Red Cross in the 2015 World Disaster Report, and the United States Federal Communications Commission in 2016. Unfortunately, although the use of open data is announced and promoted, there is a limited insight on whether the data collected can be freely re-used by third parties.

This example, however, is not the only one, since open data became widely used in this category and Haiti and Philippines are two more examples, which used open data along with the aid statistics to provide a response to disasters in a similar manner.

### 3.3 Healthcare area

The Healthcare sector is one of the most important sectors, which became even more important in the light of the COVID-19 pandemics. However, while the role of open (government) data during the COVID-19 pandemic was widely discussed and the open data enabled countries to track the current state of the art and spread of the disease, as well as analyse the trends of its spread etc., and triggered the development of various related services such as dynamic visualizations, contact tracking apps, mobility-related apps and modelling the spread of the disease for better informed decision making on the restrictions to be made (López et al., 2021) etc., the popularity and the importance of more traditional healthcare-related open data remain the same. Here, the access to personalized healthcare providers data utilizing an open-data-based platform for comparison and evidence-based decision-making come, where DATA ATuServicio in Uruguay, with which we are familiar with, is an example we would like to refer to.

*Challenge and Value*

The web-based Atuservicio.uy application was developed by DATA - Open Data, Transparency and Access to Information (DATA is the Spanish abbreviation), a civic technology, non-government organization, and the Ministry of Health, supported by ILDA as part of its strategic initiatives programme. It relies on the principles of open government, open data, transparency, freedom of information (FOI) and participation through the use of civic technology (Scrollini, 2017). This application provides citizens with an opportunity to access personalized healthcare providers data utilizing an open-data-based platform for comparison and evidence-based decision-making on which health service provider should be given a preference at a critical time, i.e., when the choice on a health service provider should be made - this choice should be done by Uruguayans every February (Scrollini, 2016). It is considered that this application works on creating social tools to promote participation and public debate through transparency, open data, and access to information.

According to the previous study of one of us (see (Flores, 2020)), it focuses on open government principles through the use of civic technology, which can be described by three aspects: (1) the creation of social tools for participation and the reuse of open data, in collaboration with their partners and the community; (2) the strengthening of a local and regional civic tech community, organizing networks and events in Uruguay and abroad; and (3) social activism, working as part of a network promoting open government, open data, transparency, freedom of information and participation.

On their website they claim that they are a horizontally managed and consensus-based organization, emphasizing that most of their projects have been co-created either with experts on this subject (Government, Academia, Media, other CSOs) or with the community. DATA played a significant role in advancing open data-related initiatives in Montevideo, and the city has benefited from its presence (Bonina and Eaton, 2020). Furthermore, the Latin American Open Data Initiative (ILDA) stated that the tools developed by the organization also generate essential contributions to the open data use and open ecosystem in two ways (ILDA, n.d.): (1) in the reusing and publishing of open datasets from each of their projects; and (2) in the highlighting of the need for improving data quality and availability.

According to (Scrollini, 2016), more than 35 000 Uruguayans accessed this service once it has been launched and find it valuable, while in the second year, its audience increased to 60 000 users. Sangokoya et al (2015), however, have managed to determine the impact of the Atuservicio.uy on its intended beneficiaries, where in addition to the "average citizens", several more beneficiaries were identified - health providers, government agencies, the media and civil society and unions and a synthesis of this impact was provided. They found that this service enables "average citizens" to make better informed health decisions, provide them with data-driven evidence and tools to make better decisions on the healthcare choice, as well as promoting their activities as a facilitator of monitoring and continuously assessing the health service they received. Health providers, however, can make it clear to citizens which health options are most suited to their needs and improve the quality of responsiveness of services based on data-driven demand from citizens, while government agencies can improve the public health system with greater transparency, efficiency, and accountability. The media is able to promote better data journalism efforts and data-based arguments for public debate on healthcare, while civil society and unions are able to better inform argumentation and advocacy of the healthcare system.

## 3.4 Cities and Regions area

According to the classification we utilize, this category refers to the services, which aim at improved data sharing on energy, transportation, water, waste, human traffic, etc., community decentralization in suburban and rural areas, thereby respecting the diversity. This means that actually this category overlaps some other categories, such as the energy-related one. Therefore, we will not refer to the above discussed examples and, instead, will cover other showcases representing one of the most discussed in the literature - transportation-related showcases of Finnish Transport and Transport for London. We will conclude with a less mature example fitting this category emphasizing that even very limited resources may launch great SDG-compliant initiatives.

*Challenge and Value*

First, let us refer to the case of Finnish Transport (Leviäkangas and Molarius, 2020), which estimated annual increased marginal turnover enabled by the open government data (OGD) of the private companies yielded to a minimum of 102 million EUR, while the estimated annual gross value added to the economy based on the use of OGD was 41 million EUR. They refer to the case of Finnish Transport Safety Agency's (Trafi) data, which allows them to demonstrate how OGD are utilized by business actors and how they are capable of being translated into economic gross value added. They found that the largest group of benefitting companies was the insurance and financial services, which stated that open data from Trafi generated 'tens of millions of indirect business benefits. They were followed by marketing and publishing. In addition, the open data was seen as an essential business asset by companies which developed their business and services based on the data and thus, created jobs and economic well-being for the society.

Probably the most well-known example with impressive quantitative assessment is the British national use-case of one of the world's largest public transport operations, Transport for London (TfL). TfL have managed to transform the availability of real-time data of both live arrivals, timetables, air quality, network performance and accessibility, for its customers and personnel, using the open data approach, and what the results of this transformation were thereby estimating the benefits that public transport authorities could gain from the publishing of data as open data (Stone and Aravopoulou, 2018)). The provided example of the data on the use of transport for London allowed them to prove both qualitatively and quantitatively that the benefits of data opening can be as high as those of major transport infrastructure development projects. The savings to individual passengers and businesses from opening TfL's data were quantified in £130m per year. They emphasized an importance of defining and implementing an open data approach, stressing an importance of a clear commitment that data belong to the public and that third parties should be allowed to use and re-use data, by having a strong digital strategy, and through creating a strong partnership with data management organizations that can support the provision of large amounts of data.

Data opening allowed passengers and other road users to gain a better travel experience. This led them to the preposition that open data approach can be assessed as a financial/ economic contribution to customers and organizations, thereby contributing significantly to the society and improving their daily activities. Considering the specificity of the open data and their availability for both commercial and non-commercial purposes, the data are also used by external companies, which in this case were Waze, Twitter, Google, Apple, Citymapper, Bus Checker, Bus Times and Mapway, as well as academics and professional developers. They found that such a wide availability and accessibility of data can lead to creation of more than 600 customer-related products and services and more than 12 000 registered developers, whose number increased more than 10 times in the last 7 years, reaching millions of active users, enabling passengers and other road users to gain a better travel experience. However, it should be noted that although more than 600 services have been developed, same as with open data, not all of them are used and only 40 of those services were recognized as popular and used. Most of them are travel apps, which is not a surprise. But what is interesting, is savings, i.e., the authors found that live bus travel data accessed via these apps saves £58m a year, helping Londoners plan better routes and avoid long waits at bus stops. The authors highlight that interest in real-time open data has produced a feedback loop that has encouraged all data contributors within Transport for London to improve their data in terms of both granularity and quality, which should be considered crucial for data re-use and success of a third-party app ecosystem. Similarly, this movement has led to the improvements for the portal since in July 2010 due to a popularity of the data feed, which caused a temporary halt to be called on the newly introduced API feed because of huge demand by apps that use the service.

This portal provides a wide variety of open data, including less transportation-oriented data, such as the open data on the air and its quality provided by the Kings College London in the form of London Air API. It provides users with the up-to-date data on air quality in London. This, however, points to another very popular movement of open-air data. This example is even more interesting in the light of some cases, where although the gathering, collection and supply of these data to the public may not be financially supported at a governmental level, being of high interest and value for citizens, they tend to be collected and provided in a more collaborative nature. This is the case for Latvia and PilsētaCIlvēkiem (City for People, https://gaiss.pilsetacilvekiem.lv/) project standing for "open data air quality monitoring network", where the organizers of the project encourage citizens to join this movement and built their own sensor (according to the provided instruction or by ordering already ready sensor) for the collection of the respective data and distributing them as open data. At the moment this resulted in the establishing of 34 sensors in the capital of

Latvia, while the intended number is of 100 sensors. Although this portal does not follow the best practices facilitating collaboration and cooperation by involving multiple stakeholders, this still became popular among citizens.

At the same time, there are significantly more successful examples such as the Netherlands and Amsterdam, the USA and Chicago, Germany and Darmstadt, and Taiwan and Taipei. The last example (Chen et l., 2017), however is known for introducing an open participatory urban sensing framework for PM2.5 particle monitoring, which then, in collaboration with maker communities, industry partners, and the Taipei City Government, has managed to develop various types of devices for different communities, which by May 2017 were deployed not only in Taiwan, but also in 29 other countries. The data collected from these devices are then deployed in the form of the real-time open data, which allowed them to become one of the largest deployment projects for PM2.5 monitoring in the world, making them available to all stakeholders, including academic organizations and research groups, thereby allowing them to conduct research on air pollution along with any associated respiratory diseases, i.e. PM2.5 is directly related to many serious health problems, such as lung cancer, premature death, asthma, and cardiovascular and respiratory diseases. To sum up, this project contributes to the development of new services using data made open, thereby creating a chain of valuable open data-based solutions and services.

## 3.5   Finance area

Another sector to be discussed is finances, where the philosophy of the Society 5.0 expects the diversification of financial services with the help of digital technologies, which should lead to the improved distribution of funds across society and improved access to financial services based on utilization of crypto-currencies and token economies.

Although the very direct connection between the financial or banking sector and the open data could be not very obvious, rather than speculating on the positive consequences and effect of opening and reusing the open data, there are known real examples, where these concepts are very interconnected. More precisely, the concept of open data triggered the development of a completely new and probably even revolutionary step in this area called "open banking", considered to be a powerful tool for the FinTech industry. While this concept is relatively new for many countries, some leaders in areas of digital literacy and open data, have already implemented this initiative and actively work on the increase of its popularity, although there are known positive examples in less developed countries as well.

*Challenge and Value*

According to (Open Banking, 2021), Open Banking "uses secure application programming interface (API) integration with banking systems to let consumers share their banking data with third-party fintech providers for new and innovative financial products and services", where Nigeria took the leadership for this movement in Africa. This should allow equality in rules and principles, data sharing and access to them by actors of the banking ecosystem, by means of its standardization and opening of the banking ecosystem to external developers, thereby eliminating the gap between companies of different sizes or maturity levels and allowing better personalized services, i.e., fitting better their needs. Another closely linked concept is open finance (sometimes used as synonyms), where any type of financial data may be shared on the informed consent basis from any type of business. This particular sector requires cooperation and collaboration of many stakeholders as well, since the success of the open baking initiative depends on the stakeholders.

However, probably the best known and mature example in this area is the BBVA API market in Spain. It is expected that the paradigm of openness and open banking, in particular, will enable generation of new models based on a richer set of data to be used for this and respective changes in the whole banking ecosystem. At the same time, this should encourage creation of new alliances and scenarios of partnership at both local and global scales, which should be also more efficient since they will work on the basis of the APIs intended to be used despite the industry and business, consequently providing better service, which should better meet users' needs and expectations. The creation of both individual models and the whole ecosystem is intended to be shared thereby eliminating inequality between different countries or enterprises.

There are two notes to be added to this. Firstly, although the role of openness as a philosophy is at the core of this showcase, the open data used and produced by these services do not comply with the concept of the open data in a full manner as can be understood from the above. Second, although we do not intend to discuss this in very detail, it should be emphasized that this particular area requires extensive research on the legal, trust, privacy etc. issues, which is also something to be described as a new opportunity brought by this transformation but also a potential barrier and determinant of the resistance to this invention.

### 3.6 Agriculture and Food area

Another sector that is now popular in terms of both the SDGs and open (government) data (with the particular popularity in Mediterranean countries), is agriculture and food, including, but not limited to the use of crop growth technology, food safety (see also the European Food Safety Authority (EFSA)) and the optimization of the food value chain.

It indisputably requires an inclusion of various actors as a prerequisite for the success. Here examples deserving the attention are Global Open Data for Agriculture and Nutrition and global food security programme Plantwise.

*Challenge and Value*

The idea of the Global Open Data for Agriculture and Nutrition (GODAN, https://www.godan.info/) is to make open data on agriculture and nutrition available, accessible, unrestricted and usable, i.e. open data principles-compliant, for stakeholders that should make it possible to tackle extreme poverty, eradicate hunger, improve nutrition, and achieves food security. GODAN together with their partners aim to build high level policy and private sector support for open data, thereby enriching both the current set of data they provide and the variety of further re-uses of these data.

In the light of the required inclusion of different actors, GODAN network supports and encourages collaboration and cooperation across existing agriculture, nutrition and open data activities and stakeholders to identify and then resolve long-standing global topic-related challenges.

As one of examples, they mention the challenge related to the need to support agricultural innovations in developing countries such as West African countries, where the cocoa plantations are not as productive as expected, leading to low income for farmers, and given other important factors affecting the progress, including but not limited to individual rights and the effects of climate change, poor soil fertility, the knowledge gap on good crop nutrition and proper management of cocoa trees etc. The solution provided to respond to this challenge is called CocoaSoils (https://cocoasoils.org/) - project, which establishes long-term trials on cocoa, fertilization, and production. and deploys an Integrated Soil Fertility Management (ISFM) approach embracing an integrated system's approach to cocoa intensification while combining improved planting materials, canopy cover management and pest/disease control with targeted fertilizer application to enhance sustainability and avoid deforestation. This is complemented by (re)-use of locally available (organic) nutrient sources and appropriate intercropping. It works across the globe, with trials in six countries working with over 90,000 farmers while also considering aspects of deforestation, child labour and ecosystem services. The partnership is led by the International Institute of Tropical Agriculture (IITA), WUR and IDH, and involves a range of companies: Mondelez, Nestle, Barry Callebaut, Mars, Yara, ICL, and research institutes: CIAT, UN WCMC, and institutes for cocoa in Ghana and Cote d'Ivoire. As a result, the GODAN advocacy for open data, the partnership leadership has committed to publish all data as fully open (CC-BY) upon their availability. While the data of the current project is the first step, it is envisioned that other data sources on cocoa can also be captured as part of the project. They emphasize that benefits from open data can be gained by both commercial organizations, governments, farmers, traders and value-added resellers, science community, consumers, stressing that the open data integration with methods and practices with advocacy, technical support and senior policy guidance, are capable to improve not only the food security and nutrition but also to help ensuring zero hunger and empowering the lives and livelihoods of people and farming communities across the globe (Worldbank, 2022).

This opinion is also shared by another example - Plantwise Global food security programme (https://www.plantwise.org/impact/plantwise/) led by CABI, which intends to increase food security and improve rural livelihoods by reducing crop losses. Weak knowledge on the food security and complexity of making decisions on the diagnoses and recommendations regarding food safety, related or implying challenges, lead the developers to the idea of creating a global knowledge resource for plant health information called the PlantwiseKnowledge Bank, which combines global and local open data, providing sufficient information for the experts to turn questions into answers. This provides doctors with the right diagnoses and recommendations. This, however, allowed the establishment of sustainable networks of more than 5000 plant clinics (compared to 1800 in 2016), run by trained 13200 plant doctors (compared to 5000 in 2016), where more than 54 000 000 farmers (compared to 4500000 in 2016) receive practical plant health advice. Working in close partnership with over 170 in-country partners, Plantwise strengthened national plant health systems from within, enabling countries to provide farmers with the knowledge they need to lose less of what they grow.

They also support and encourage cooperation and collaboration of multiple stakeholders, in terms of both different countries and respective government (at least some of them are - Australian Centre for International Agricultural Research, Ministry of Agriculture and Rural Affairs (MARA, People's Republic of China), Ministry of Foreign Affairs of the Netherlands, Swiss Agency for Development and Cooperation SDC, UK aid from British people and European Commission), as well as individual participants such as researchers, thereby building more comprehensive base of knowledge and digital skills to strengthen plant health systems, increase the supply of higher quality and safer food, improve availability of safer plant protection products, strengthen detection and response to pest outbreaks.

## 3.7 Public service area

Another sector representing public services is seeking for improved services by public administration based on digitalization and improved data sharing that should result in establishing safety nets in response to safety issues. This is something overlapping with showcases we have elaborated on above in the context of disaster prevention and healthcare, although there can be provided other examples as well.

*Challenge and Value*

The example connecting not only citizens of two countries but rather countries of two continents is Papelea, which provides the support for citizens of Spain and Mexico and solves their questions of both legal and administrative nature by collecting public information from various websites of these countries and providing it in a summarized and structured manner. In addition, it connects users with professionals in each field, so that they can answer their questions and offer them their services. It uses open data rather than the secondary data source, although other materials used are freely and publicly available information, which should not necessarily fulfil the open data principle but be compliant with basic principles of openness.

Similar approach is seen in the web portal of Buenos Aires titled BA Obras (BA Public Works), where the most up-to-date information on public works, including but not limited to works' budget, start and finish dates, contractor and bidding documents, and preview progress with videos and photos, and other information, which can be useful for citizens is provided mainly by means of visualization, thereby reducing the level of complexity of provision of this information (Straface et al., 2019). The platform utilizes the user-centred design to allow the increasing use by residents, NGOs and journalists. This makes it a tool to engage locals in participatory processes such as town hall meetings, or co-creation urban design processes.

Another interesting and impressive platform is Codeando Mexico (http://codeandomexico.org/) with a very strong encouragement of collaboration and community engagement for improving the city and its citizens by providing both the support and knowledge and skills (if needed) and opportunity to change the city by either reporting on the ongoing issue or challenge to be addressed or getting involved in the resolving already identified being in line with the Smart City and Smart Society paradigms and facilitating their citizen-centred continuous improvement.

### 3.8 Logistics area

Another sector expects the combination of the concepts we consider, which in combination with technology would facilitate automation of logistics by means of data sharing across the whole supply chain.

*Challenge and Value*

The example we would like to draw the attention here is compliant with the topics covered and the concept of Industry 4.0 with the step towards Industry 5.0 and technologies expected to be used by the Society 5.0 - FIWARE-based automated logistics system developed by the University and the RoboticBase. It uses heterogeneous mobile robots that enable the management of real-time data from various robots, IoT sensors and open data (by means of Open Data Management Systems (ODMS) web application called Idra) to control and manage robots and their operations. It is expected to be widely used and be value-adding in multiple "smart areas" such as smart city, smart water, smart industry and smart agri-food, which makes this ongoing project a perfect example of combination of these concepts.

The general idea of the whole project is also compliant with the open science principles and openness as a philosophy, including a strong support of community engagement.

And the last area defined is related to manufacturing and services where the focus is switched from the hardware to services, with the prerequisite of providing the customers with an opportunity to get items specifically designed for their needs, as well as providing a support for small business. Respective examples have been already discussed above and demonstrated the role and result of the use of open data clearly (e.g., financial sector and "open financing" and "open banking" cases discussed above), therefore we will not discuss this area separately. However, the point we would like to emphasize is that in addition to open data, the paradigm of openness is the key to success in this case.

### 3.9 Summary

Now, let us discuss whether there exists a relationship between the open data and features of Society 5.0. Table II provides a matrix of focus areas, which use-cases have been discussed above, and Society 5.0 main features. This very simple matrix makes it obvious that the concepts of the Society 5.0 and open data-driven services are interrelated and share similar objectives, although in some cases these links are stronger compared to others.

As an example, services using the open data as a primary data, i.e., open data-driven services, and those using open data as a supplementary data or producing the open data as a result of their operation, are focused on *problem solving and value creation*, with prevailing number of those intended for "*a society, where value is created*", where cooperation, collaboration, other parties and community engagement become crucial. In some areas covered, i.e., disaster prevention, public services and healthcare, this form of the "society" becomes a key for success. PDCA cycle and strict set of guidelines to be followed is not mandatory anymore and more freedom is given, including but not limited to the increasing popularity of the co-creation. However, it should be acknowledged that some showcases, opposite to the abovementioned, seeks for standardized approaches as was the case for Finance sector showcases and is sometimes the case for Agriculture and Food, since the complete avoidance of the unification and standardization of processes in many areas will lead to fail instead of the promised success as it could be expected from this Society 5.0 principle, if treated in a too straightforward manner and extrapolated on all cases and areas.

These services also support and promotes *diversity*, where not only a society but also different types of organization despite their size and finances are expected to be provided with the same set of data, thereby eliminating discrimination and alienation. This intends to give businesses the opportunity to provide their unique and diverse services without the need to collect and process data, which may turn into a resource-consuming task by re-using the data already collected and ready for the value extraction from them, thereby allowing them to identify diverse tasks, challenges and needs and turn them into real business. Moreover, as regards the determination of these tasks, it is expected to be done not only by users or businesses but also by the whole society as is expected to be done in the area of Public Services, where the community engagement supposes notifying the audience (not only the portal holders) about the current challenges existing and requiring actions.

This is also the case for *decentralization*, which is in the core of the openness and open data - data being available and accessible to anyone and anywhere, thereby liberating the society from disparity. This remains to be the focus for the services created using open data since in many cases their aim is to provide the service to be used and be beneficial by the whole society. Here, the cases belonging to Disaster Prevention, Healthcare and Public Service, Agriculture and Food, Finance, Cities and Region are the most expressive examples due to their target audience.

*Resilience* and sustainable development aimed at ensuring liberation from anxiety and vulnerability from disasters, terrorism and attacks in cyberspace is an interesting point to be mentioned. Although on the one hand the open data and respective reuses in the covered focus areas intend to ensure resilience, some of these advances as is the case for the financial sector, for instance, bring new challenges regarding safety and security of users and their data - although the general aim is the same, the related issues to be addressed are controversial. Therefore, it is questionable whether this principle can be considered as fully fulfilled and sharing the same objectives in both cases. On the other hand, in other areas such as Disaster Prevention, this objective is completely compliant with ensuring liberation from vulnerability from disasters, while Healthcare, Agriculture and Food, and Public Service definitely complement the collection of the respective examples.

And last but not the least is *sustainability and environmental harmony* with the aim of transforming the current form of the society to a society, where humankind lives in harmony with nature with the liberation of resources and environmental constraints and mass consumption of resources with a major impact on the environment, where the Energy area is probably the most accurate example of sharing this objective. Otherwise, this principle is sometimes understood as a long term objective for open data-driven services, particularly for those oriented to Smart Cities, Smart Living etc., with less focus on environmental harmony in a general case.

**Table 3.1**. Summary of use-cases by focus area

| Use case by focus area / principle of Society 5.0 | problem solving and value creation | diversity | decentralization | resilience | sustainability and environmental harmony |
|---|---|---|---|---|---|
| Energy | + | ++ | ++ | ++ | ++ |
| Disaster Prevention | ++ | + | ++ | ++ | ++ |
| Healthcare | ++ | + | ++ | ++ | + |
| Cities and regions | ++ | + | ++ | ++ | ++ |
| Agriculture and food | ++ | + | ++ | ++ | ++ |
| Logistics | + | ++ | ++ | + | + |
| Public services | ++ | ++ | ++ | ++ | + |
| Finance | ++ | ++ | ++ | + | + |
| Manufacturing and services | + | ++ | ++ | + | + |

To sum up, the open data by themselves, i.e., by definition, fulfil such requirements of the Society 5.0, as diversity and decentralization, while problem solving and value creation, as well as resilience, are objectives to be met by the reuse of open data - their transformation into the value facilitating innovation and promoting societal progress with slightly different definition of sustainability and environmental harmony.

# 4 DISCUSSION AND CONCLUSIONS

To sum up, although there are relationships between the concepts of open (government) data and Society 5.0, there are some significant differences, implying the main aims of both. In other words, they share the same philosophy of openness, equal rights, diversity etc., but Society 5.0 can be understood as more isolated and society-oriented with transformation mainly affecting the way of thinking and living. Open data, in addition to being a driver for "collective intelligence" also known as "wisdom-of-crowd", is a resource and tool of transformation of not only the society but also all areas of people's lives, including the forms of governance and response to both daily challenges and different types of crises, including but not limited to economic, natural crises and those requiring a response in a short time. However, the question to be asked here is "Are the open (government) data the answer?".

We believe that the open (government) data (innovation / services) trend develops a new form of society, which we refer to as "open data-driven society", which forms a bridge from Society 4.0 to Society 5.0. However, although it could be speculated that the presence of OGD portal and availability of open data together with the active citizens, who would like to contribute and get involved in some activities, is the answer to all the issues within the country, unfortunately, it should be admitted that it is not a "silver bullet". This is all the more case, given that achieving sustainability is a challenging task, which is only possible through sociotechnical transition, i.e., through the adoption of technological innovations in a complex social system with multiple parties / stakeholders (Kroh, 2021). In other words, not only users' imagination and the desire to create value from the OGD for society do matter, but also the diversity of the open (government) data and the readiness of the country under question for digitization at all levels.

In every case presented, the availability of open (government) data, obviously compliant with open data principles, and the knowledge of open (government) data, was a precondition for the creation of the service. Whereas before the emphasis was on the government to release data, the mental model has now switched, and it is important not only to release data in an open format, but also to encourage to use them and create new services from them, perhaps in a co-creative manner, i.e., ensuring an effective multiple stakeholder cooperation or collaboration. Thus, open data-driven co-created services do not only represent a new technological concept, but also represents a paradigm shift in the way of thinking and collaboration of different levels (individuals, public authorities, government), while the changes in the behaviour of data users lead to the development of the Society 5.0. The later point, however, once again brings us to the issue of digital literacy since the data are mainly used and transformed into the service or tool when the prospective users are capable of doing so, i.e., they have appropriate knowledge and skills of using data. There might be some exceptions, when open data portals have some data transformation in-built tools or educational materials, as well as when workshops and / or hackathons are organized to give this knowledge, but, unfortunately, it is still not very widely spread practice, although attempts are seen very often.

Not least important point is data quality and data credibility, which became more than ever topical and common for many countries in terms of pandemics. This once again proves that although data availability is the prerequisite, a shift should be made from "quantity to quality". While data quality can be examined to some extent, however, it is not always clear for the OGD who is responsible for this (if any) or the data user should carry out data quality analysis prior to their use, data credibility is a more complex issue. This is also the case for (Yiannakoulias et al., 2020), which points to inaccuracies in the OGD pandemic data with low short-term decision-making value, encouraging data publishers to highlight the uncertainties associated with the data, because while the OGD can increase the government transparency and accountability, it is essential that all publication, use and re-use of these data highlight their weaknesses to ensure that the public is properly informed about the uncertainty associated with these data and their reuse will bring added value, which is only possible if data are correct, reliable and trustable.

What is especially important in the light of active open data reuse and co-creation in particularly, is that the more stakeholders are involved especially those who are ready to transform data into the solution or service, the more qualitative data may potentially become since reusers are capable to notice inaccuracies and mistakes in the data published. Then, if the communication and / or feedback mechanisms are in place, the communication with the data publisher can take place to inform them about this issue with hopefully data improvement. We knowingly emphasize here those users who are more likely to become reusers – this is due to the fact that typical users, who are less likely to constantly use the data, are more under the risk of losing the trust and interest in the data, if the quality issues are identified. Here, definitely "quality-by-design" is a more appropriate paradigm, which, at the moment, is far away from the reality.

The COVID-19 driven demand for actually important / high-value and timely data increased the reuse of OGD. Although the most popular and required was the health category, the reuse of other dataset categories also contributed to raising the awareness about the importance of OGD for crisis management through co-creation of web and mobile applications. On the other hand, if the OGD ecosystems are not prepared for this increased demand, whether in terms of datasets quality or existence of a robust and reliable data infrastructure, their change is inevitable. If clear procedures are not defined and OGD standards are not implemented properly, new approaches and channels for communication and information delivery emerge in the OGD ecosystem. They affect and reshape the relations and interactions between stakeholders and other components. Taking a step back to the necessity for timely and up-to-date data, we should mention that this is a key for data reuse to make the service to be created based on these data to be of interest to the audience and make it dynamic and as close to addressing the real-time situation as possible. Here, in addition to data availability and timeliness, compliance with such principles as machine-readability, data currency, which suppose data to be regularly updated, as well as retrievability by means of data accessibility via API support becomes almost a prerequisite.

The range of stakeholders' skills can also limit the success of OGD ecosystems. Of course, not of least importance is the overall digital literacy of the society and the fact of the resistance to the use of open data, which can be extrapolated to other areas (digitization, transformation in the Society 5.0 etc). This seems to be important in terms of both, the developers of services and the interest of the society and more important their ability to use these services. This is something similar to what we already covered in section 3, where in the case of Transport for London, less than 7% of services are used. This is a question, which is often covered in studies devoted to the technology acceptance and determination of their determinants by means of UTAUT (the Unified Theory of Acceptance and Use of Technology), TAM (Technology Acceptance model), TOE (Technology-Organization-Environment framework) etc. or those, exploring innovation resistance of both the passive and active forms, using one of relevant models such as Innovation Resistance Theory (IRT).

In other words, there is a significant room for improvement at levels of both open data ecosystems and society, which limits the potential of open data.

However, even in the light of these challenges, which are not likely to be quickly resolved, the trend of openness at all levels with a particular focus on open data develops a new form of society we call "open data-driven society", which forms a bridge from Society 4.0 to Society 5.0. What is more, given that Society 5.0 aims to address and resolve the critical social issues facing humankind today, which refers to the categories of open data found to be popular and highly valuable today, we find it important to conduct future studies addressing the link between the open (government) data and trends in terms of their topic coverage in relation to the Sustainable Development Goals. Although we have partly covered the existence of these relationships here, more in-depth analysis should be made by raising awareness on this and defining further actions to be made to achieve SDGs by means of the OGD as is done actively by several countries such as Spain and respective Spain open data portal. This should allow better identification of the role of openness in promoting human-centric Smart Society, Smart city, and Smart Living.